\documentclass{article}
\usepackage{spconf,amsmath,graphicx}
\usepackage{textcomp}
\usepackage{psfrag}
\usepackage[printonlyused]{acronym}
\usepackage{lipsum}
\usepackage{booktabs} %nice tables
\usepackage{multirow}
\usepackage[tight]{subfigure}
\usepackage{pgfplots}
\usetikzlibrary{patterns}
\usepgfplotslibrary{units}

\pgfplotsset{compat=1.3}
\pgfplotsset{
  /pgfplots/xbar legend/.style={
  /pgfplots/legend image code/.code={%
  \draw[##1,/tikz/.cd, bar width=3pt,yshift=-0.1em,bar shift=0pt]
  plot coordinates {(0.8em, 0cm) (0.6em, 1.7*\pgfplotbarwidth)};},
  }
}

\acrodef{ASR}{Automatic Speech Recognition}
\acrodef{CDR}{Coherent-to-Diffuse Power Ratio}
\acrodef{CHiME-3}{3rd CHiME Speech Separation and Recognition Challenge}
\acrodef{DNN}{Deep Neural Network}
\acrodef{DoA}{Direction-of-Arrival}
\acrodef{FIR}{Finite Impulse Response}
\acrodef{GMM}{Gaussian Mixture Model}
\acrodef{GPU}{Graphics Processing Unit}
\acrodef{HMM}{Hidden Markov Model}
\acrodef{LS}{Least-Squares}
\acrodef{MFCC}{Mel-Frequency Cepstral Coefficients}
\acrodef{MMSE}{Minimum Mean Square Error}
\acrodef{MVDR}{Minimum Variance Distortionless Response}
\acrodef{PDF}{Probability Density Function}
\acrodef{SNR}{Signal-to-Noise Ratio}
\acrodef{STFT}{short-time Fourier transform}
\acrodef{STFT-UP}{STFT uncertainty propagation}
\acrodef{TDOA}{Time Difference of Arrival}
\acrodef{WER}{word error rate}
\acrodef{WNG}{White Noise Gain}
\newcommand{\bb}[1]{\mathbf{#1}}

\newcommand{\sine}[1]{\ensuremath{\mathrm{sin}(#1)}}
\newcommand{\cosine}[1]{\ensuremath{\mathrm{cos}(#1)}}
\newcommand{\sinc}[1]{\ensuremath{\mathrm{sinc}(#1)}}
\DeclareMathOperator*{\argmin}{argmin}
\renewcommand\Re{\operatorname{Re}}

\title{Robust coherence-based spectral enhancement for distant speech recognition}
\name{\parbox{0.9\textwidth}{\centering Hendrik Barfuss, Christian Huemmer, Andreas Schwarz, and Walter Kellermann\thanks{The research leading to these results has received funding from the European Union's Seventh Framework Programme (FP7/2007-2013) under grant agreement n$^\mathsf{o}$ 609465 and from the Deutsche Forschungsgemeinschaft (DFG) under contract number KE 890/4-2.}}}
\address{Multimedia Communications and Signal Processing,\\
         Friedrich-Alexander University Erlangen-N\"urnberg\\
         Cauerstr. 7, 91058 Erlangen, Germany \\
         {\{barfuss,huemmer,schwarz,wk\}@lnt.de}}

\begin{document}
%\ninept
%
\maketitle
\begin{abstract}
%The abstract should contain about 100 to 150
In this contribution to the \ac{CHiME-3} we extend the acoustic front-end of the \ac{CHiME-3} baseline speech recognition system by a coherence-based Wiener filter which is applied to the output signal of the baseline beamformer. To compute the time- and frequency-dependent postfilter gains the ratio between direct and diffuse signal components at the output of the baseline beamformer is estimated and used as approximation of the short-time signal-to-noise ratio. 
The proposed spectral enhancement technique is evaluated with respect to word error rates of the \ac{CHiME-3} challenge baseline speech recognition system using real speech recorded in public environments. 
Results confirm the effectiveness of the coherence-based postfilter when integrated into the front-end signal enhancement.
\end{abstract}

%reset acronyms
\acresetall

%--------------------------------------------------------
% Introduction
%--------------------------------------------------------
\section{Introduction}
\label{sec:intro}

%-----------------------------
% Robust ASR in noisy environments
For a satisfying user experience of human-machine interfaces it is crucial to ensure a high accuracy in automatically recognizing the user's speech. As soon as no close-talking microphone is used, the recognition accuracy suffers from reverberation as well as background noise and active interfering speakers picked up by the microphones in addition to the desired speech signal~\cite{delcroix_2013,Yoshioka_2015}. Signal processing techniques for robust speech recognition in noisy environments can be categorized into two major categories, namely front-end (e.g., speech enhancement \cite{cohen_2003,krueger_2010,gales_2011}) and back-end (e.g., acoustic-model adaptation \cite{li_2006,liao_2013,yu_2013}) processing techniques.

%CHiME challenge
The \ac{CHiME-3} \cite{Barker_Chime3} targets the performance of state-of-the-art \ac{ASR} systems in real-world scenarios. In this year's challenge, the primary goal is to improve the \ac{ASR} performance of real recorded speech of a person talking to a tablet device in realistic noisy environments by employing front-end and/or back-end signal processing techniques.

%-----------------------------
% Overview of signal enhancement with back-end from baseline
In this contribution to the \ac{CHiME-3} challenge, we focus on front-end speech enhancement and extend the \ac{CHiME-3} baseline front-end signal processing, consisting of a \ac{MVDR} beamformer, by a coherence-based postfilter. The postfilter is realized as a Wiener filter, where an estimate of the ratio between direct and diffuse signal components at the output of the baseline \ac{MVDR} beamformer are used as an approximation of the short-time \ac{SNR} to compute the time- and frequency-dependent postfilter gains. The employed postfilter is \ac{DoA}-independent and has a low computational complexity.

\begin{figure*}[t]
  \centering
  \psfrag{STFT}[c][c]{STFT}
  \psfrag{BF}[c][c]{\parbox{2cm}{\centering Beamformer\\ (Section~\ref{subsec:MVDR})}}
  \psfrag{PF}[c][c]{\parbox{2cm}{\centering Postfilter\\ (Section~\ref{subsec:CDRpostfilter})}}
  \psfrag{FeatExTr}[c][c]{\parbox{2cm}{\centering Feature\\ extraction/ \\transform.}}
  \psfrag{DNN}[c][c]{\parbox{2.4cm}{ \centering HMM-GMM/ \\ HMM-DNN}}
  \psfrag{Rec}[c][c]{\parbox{2cm}{\centering Recognition}}
  \psfrag{Base}[c][c]{Baseline acoustic back-end system (Section \ref{sec:BackEnd})}
  \includegraphics[scale=.7]{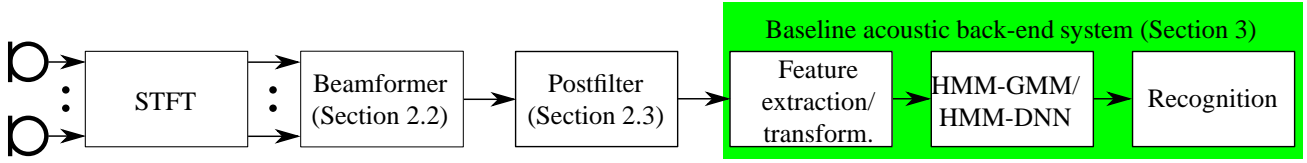}
  \caption{Overview of the overall signal processing pipeline system with beamformer and postfilter as acoustic front-end signal processing. The acoustic back-end system, including feature extraction/transformation, is equal to the baseline acoustic back-end system provided by \ac{CHiME-3} \cite{Barker_Chime3}.}
  \label{fig:ASR_pipe}
\end{figure*}

%Overall signal processing
An overview of the overall signal processing pipeline is given in Fig.~\ref{fig:ASR_pipe}. Whereas the purpose of the beamformer is to reduce the signal components from interfering point sources by spatial filtering, the postfilter shall remove diffuse interference components, e.g., reverberation, from the beamformer output signal. The output of the front-end signal enhancement (consisting of \ac{MVDR} beamformer and postfilter) is further processed by feature extraction/transformation and acoustic modeling following the \ac{CHiME-3} baseline \ac{ASR} system, which provides a \ac{HMM}-\ac{GMM}-based as well as an \ac{HMM}-\ac{DNN}-based speech recognizer \cite{Barker_Chime3}. 
%TODO: front-end highlighten

%-----------------------------
% Overview of the DNN-HMM hybrid system
The remainder of this article is structured as follows: In Section~\ref{sec:FrontEnd}, the proposed front-end signal enhancement is introduced in detail, followed by a brief review of the employed \ac{ASR} system in Section~\ref{sec:BackEnd}. The performance of the front-end speech enhancement is evaluated with respect to \acp{WER} of the baseline \ac{ASR} system, which are presented in Section~\ref{sec:Exper}. A conclusion and an outlook to future work is given in Section~\ref{sec:summary_conclusion}.

%--------------------------------------------------------
% Front-end
%--------------------------------------------------------
\section{Front-end enhancement techniques}
\label{sec:FrontEnd}
The front-end speech enhancement considered in this article consists of an \ac{MVDR} beamformer (provided by the \ac{CHiME-3} baseline) and a single-channel coherence-based postfilter. In the following, the baseline \ac{MVDR} beamformer is briefly reviewed, followed by a detailed presentation of the proposed postfilter.

%--------------------------------------------------------
% Signal model
\subsection{Signal model}
For a consistent presentation of the front-end speech enhancement considered in this work, we first introduce a signal model which will be used throughout this article.

The $N$ microphone signals of the microphone array in the \ac{STFT} domain at frame $l$ and frequency $f$ are given as:
\begin{equation}
  \bb{x}(l,f) = \bb{h}(l,f) S(l,f) + \bb{n}(l,f),
  \label{eq:signalModel}
\end{equation}
where vector
\begin{equation}
  \bb{x}(l,f) = [X_0(l,f),\, X_1(l,f),\, \ldots,\, X_{N-1}(l,f)]^{T}
  \label{eq:signalModel_X}
\end{equation}
contains the microphone signals, $S(l,f)$ denotes the clean source signal, and 
$\bb{n}(l,f)$ includes sensor noise as well as diffuse background noise components and is defined analogously to $\bb{x}(l,f)$ in (\ref{eq:signalModel_X}). Assuming free-field propagation of sound waves, $\bb{h}(l,f)$ represents the steering vector modeling the sound propagation between the desired source located at direction $(\phi_\mathrm{d},\theta_\mathrm{d})$ and all $N$ microphones:
\begin{equation}
  \bb{h}(l,f) = [e^{-j \mathbf{k}_\mathrm{d}^{T} \mathbf{p}_{0}},\, e^{-j \mathbf{k}_\mathrm{d}^{T} \mathbf{p}_{1}},\, \ldots,\, e^{-j \mathbf{k}_\mathrm{d}^{T} \mathbf{p}_{N-1}}]^{T},
  \label{eq:signalModel_H}
\end{equation}
where wavevector $\mathbf{k}_\mathrm{d}$ is defined as \cite{VanTrees:2004}:
\begin{equation}
  \bb{k}_\mathrm{d} = - \frac{2 \pi f}{c} [ \sine{\theta_\mathrm{d}}\cosine{\phi_\mathrm{d}},\, \sine{\theta_\mathrm{d}}\sine{\phi_\mathrm{d}},\, \cosine{\theta_\mathrm{d}} ]^{T},
  \label{eq:signalModel_waveVector}
\end{equation}
with speed of sound $c$ and operator $(\cdot)^T$ denoting the transpose of a vector or matrix. $\phi$ and $\theta$ denote azimuth and elevation angle, respectively, and are defined as in \cite{VanTrees:2004} with $(\phi,\theta)=(90^\circ,90^\circ)$ denoting broadside. Furthermore, the $n$-th microphone position in Cartesian coordinates is captured by the three-dimensional vector $\bb{p}_n, \, n \in \{0, \ldots, N-1\}$.

% 
%beamformer output 
The beamformer output $Y_\mathrm{BF}(l,f)$ is obtained by multiplying each microphone signal with a complex-valued filter weight $W_{n}(l,f)$, followed by a summation over all microphone channels:
\begin{equation}
  Y_\mathrm{BF}(l,f) = \bb{w}^{H}(l,f)\bb{x}(l,f),
\end{equation}
where 
\begin{equation}
    \bb{w}(l,f) = [W_{0}(l,f), \ldots, W_{N-1}(l,f)]^{T}
    \label{eq:BFweights_vector}
\end{equation}
contains the beamformer filter coefficients $W_{n}(l,f)$. 
% 
%postfilter output
Subsequently, the postfilter is applied to the beamformer output signal, yielding the overall output signal
\begin{equation}
  Y(l,f) = G(l,f)Y_\mathrm{BF}(l,f),
\end{equation}
where $G(l,f)$ describes the postfilter gains. After front-end signal enhancement, $Y(l,f)$ is fed into the \ac{CHiME-3} baseline acoustic back-end system \cite{Barker_Chime3}.

%-----------------------
%Equation of CDR estimator (needed in Section 2.3)
%-----------------------
\begin{figure*}
  % The spacer can be tweaked to stop underfull vboxes.
  %\vspace*{4pt}  
  % ensure that we have normalsize text
  \normalsize
  % Store the current equation number.
%   \setcounter{MYtempeqncnt}{\value{equation}}
  % Set the equation number to one less than the one
  % desired for the first equation here.
  % The value here will have to changed if equations
  % are added or removed prior to the place these
  % equations are referenced in the main text.
  \setcounter{equation}{15}
  \begin{equation}
    \widehat{\mathrm{CDR}} = \frac{\Gamma_\mathrm{n}\, \Re\{\hat\Gamma_\mathrm{x}\} -{|\hat\Gamma_\mathrm{x}|}^2 - \sqrt{\Gamma_\mathrm{n}^2\, {\Re\{\hat\Gamma_\mathrm{x}\}}^2 - \Gamma_\mathrm{n}^2\, {|\hat\Gamma_\mathrm{x}|}^2 + \Gamma_\mathrm{n}^2 - 2\, \Gamma_\mathrm{n}\, \Re\{\hat\Gamma_\mathrm{x}\} + {|\hat\Gamma_\mathrm{x}|}^2}}{{|\hat\Gamma_\mathrm{x}|}^2 - 1}
    \label{eq:CDR_estimator}
  \end{equation}
  % IEEE uses as a separator
%   \hrulefill
  % Restore the current equation number.
%   \setcounter{equation}{\value{MYtempeqncnt}}
  \setcounter{equation}{7}
\end{figure*}

%--------------------------------------------------------
% MVDR beamformer
\subsection{Minimum variance distortionless response beamformer}
\label{subsec:MVDR}

The filter weights of the \ac{MVDR} beamformer are determined such that the power of the noise components at the output of the beamformer is minimized, subject to a distortionless constraint in target look direction. Thus, the constrained optimization problem of the \ac{MVDR} beamformer is given as \cite{VanTrees:2004}
\begin{equation}
  \bb{w}_\mathrm{MVDR}(l,f) = \argmin\limits_{\bb{w}(l,f)}^{} \bb{w}^{H}(l,f) \bb{S}_\mathrm{\bb{nn}}(l,f) \bb{w}(l,f)
  \label{eq:mvdr_main}
\end{equation}
subject to
\begin{equation}
  \bb{w}^{H}(l,f)\bb{d}(f) = 1,
  \label{eq:mvdr_constraint}
\end{equation}
where $\bb{S}_\mathrm{\bb{nn}}(l,f)$ is the multichannel spatio-spectral covariance matrix of the noise components at the input of the beamformer, and vector $\bb{d}(f)$ in (\ref{eq:mvdr_constraint}) represents the steering vector corresponding to the beamformer's desired look direction $(\phi_\mathrm{d}, \theta_\mathrm{d})$, defined as
\begin{equation}
  \bb{d}(f) = [e^{-j\bb{k}^{T}_{\mathrm{d}} \bb{p}_{0} }, \ldots, e^{-j\bb{k}^{T}_{\mathrm{d}} \bb{p}_{N-1} }]^{T} = \bb{h}(l,f).
\end{equation}
Eq.~(\ref{eq:mvdr_main}) represents the minimization of the noise variance at the output of the beamformer, whereas (\ref{eq:mvdr_constraint}) contains the distortionless constraint which ensures that a plane wave coming from the desired look direction $(\theta_\mathrm{d},\phi_\mathrm{d})$ can pass the system without distortion. The optimum solution to the constrained optimization problem in (\ref{eq:mvdr_main}),(\ref{eq:mvdr_constraint}) is given as \cite{VanTrees:2004}
\begin{equation}
    \bb{w}^{H}_\mathrm{MVDR}(l,f) = \frac{ \bb{d}^{H}(f) \bb{S}^{-1}_\mathrm{\bb{nn}}(l,f) }{ \bb{d}^{H}(f) \bb{S}^{-1}_\mathrm{\bb{nn}}(l,f) \bb{d}(f) }.
    \label{eq:mvdr_solution}
\end{equation}

The multichannel spatio-spectral noise-covariance matrix $\bb{S}_\mathrm{\bb{nn}}(l,f)$ was estimated from a time interval of duration between $400\,$ms and $800\,$ms immediately before each utterance \cite{Barker_Chime3}. As in the \ac{CHiME-3} baseline, all failing microphones are excluded from the beamforming.

The \ac{DoA} was determined by using the \ac{CHiME-3} baseline localization approach which uses a nonlinear SRP-PHAT pseudo spectrum \cite{Barker_Chime3}.

%--------------------------------------------------------
%CDR-based postfilter
%--------------------------------------------------------
%TODO: wir haben in der Praxis (siehe Matlab Code) erst jedes einzelne CDR gewichtet und dann gemittelt. Im paper steht es aber genau anders herum da! -> Anpaasen und die andere Möglichkeit ausprobieren
\subsection{Coherence-based postfilter}
\label{subsec:CDRpostfilter}
As illustrated in Fig.~\ref{fig:ASR_pipe}, we apply a postfilter to remove diffuse noise components from the output of the \ac{MVDR} beamformer. The postfilter gain $G(l,f)$ at frame $l$ and frequency $f$ is given as \cite{haensler_acoustic_2004}:
\begin{equation}
  G(l,f) = \mathrm{max} \left\{ 1 - \mu \frac{1}{1 + \mathrm{SNR}(l,f)} ,\, G_\mathrm{min} \right\},
  \label{eq:WienerFilter}
\end{equation}
with overestimation factor $\mu$, and gain floor $G_\mathrm{min}$. The postfilter in (\ref{eq:WienerFilter}) is a Wiener filter using the short-time \ac{SNR} to compute the filter gains $G(l,f)$. In this work, we approximate the short-time \ac{SNR} in (\ref{eq:WienerFilter}) by the estimated \ac{CDR}, which is the ratio between direct and diffuse signal components. From (\ref{eq:WienerFilter}) it can be seen that a low \ac{CDR} value, which corresponds to strong diffuse signal components being present at the input of the system, leads to low filter gains and vice versa.

The \ac{CDR} between two omnidirectional microphones is defined as \cite{lnt2014-28}:
\begin{equation}
  \mathrm{CDR}(l,f) = \frac{\Gamma_\mathrm{n}(l,f) - \Gamma_\mathrm{x}(l,f)}{\Gamma_\mathrm{x}(l,f)-\Gamma_\mathrm{s}(l,f)},
\end{equation}
where $\Gamma_\mathrm{x}(l,f)$ is the spatial coherence function of both microphone signals. Moreover, the spatial coherence functions for the direct and diffuse sound components are given as 
%TODO: bei Gamma_n muss die Mittenfrequenz der einzelnen Teilbänder (die durch einen Index, z.B., \nu gekennzeichnet werden sollten!)
\begin{align}
  \Gamma_\mathrm{s}(l,f) &= e^{j 2 \pi f \Delta t},\\
  \Gamma_\mathrm{n}(l,f) &= \Gamma_\mathrm{diff}(f) = \sinc{2 \pi f \frac{d}{c}},
\end{align}
respectively, with \ac{TDOA} $\Delta t$ and microphone spacing $d$.

Many different \ac{CDR} estimators have been proposed in the literature, see, e.g., \cite{jeub_blind_2011,thiergart_signal--reverberant_2012,thiergart_spatial_2012}. The \ac{CDR} estimator we use in this work was proposed in \cite{lnt2014-28} and is given by (\ref{eq:CDR_estimator}), where $\Re\{\cdot\}$ and $|\cdot|$ represent the real part and magnitude of $(\cdot)$, respectively. Moreover, $\hat{\Gamma}_\mathrm{x}(l,f)$ and $\widehat{\mathrm{CDR}}(l,f)$ are the estimated coherence and \ac{CDR} of the two microphone signals, respectively. Note that $l$ and $f$ have been omitted in (\ref{eq:CDR_estimator}) for brevity.
%TODO: Wenn ich Platz brauche kann ich diese Formel laut Andreas auch weglassen
% 
As can be seen from (\ref{eq:CDR_estimator}), the employed estimator does not require the \ac{DoA} of the speech source, since $\Gamma_{s}(l,f)$ is not required for calculating $\widehat{\mathrm{CDR}}(l,f)$.
In \cite{lnt2014-28} it was shown that the employed estimator (\ref{eq:CDR_estimator}) is unbiased and robust in the sense that deviations of the coherence estimate $\hat\Gamma_\mathrm{x}(l,f)$ from the assumed model do not lead to large deviations of the \ac{CDR} estimate.
A more detailed investigation of the employed \ac{CDR} estimator (\ref{eq:CDR_estimator}) and a comparison to different estimators with respect to bias, robustness, and dereverberation performance, can be found in \cite{lnt2014-28, lnt2015-17}.
%
% 
% TODO: wenn möglich/sinnvoll noch zwei 'Kugelplots' einfügen und unbiased und robust anhand der Plots erklaeren
% 

%----------------------------------
% Description coherence-based postfilter applied to the output of a multi-channel (>2) beamformer
%TODO: begründung warum hinteres Mikrofon weggelassen? Oder diesen Kommentar weglassen_: ``excluding the backward-facing microphone,''
When applying the coherence-based postfilter to the output of a beamformer, two aspects need to be considered: First, since the microphone array of the \ac{CHiME-3} challenge consists of five forward-facing microphones, the \ac{CDR} estimator (initially designed for a pair of microphones ) has to be adapted to exploit all available microphone signals. To do so, we apply the \ac{CDR} estimator (\ref{eq:CDR_estimator}) to every pair of non-failing microphones, i.e., ten pairs for five microphones, 
%TODO: excluding the backward-facing microphone, 
to obtain the \ac{CDR} estimate of each microphone pair. 
%TODO, optional:The backward-facing microphone was excluded, since this led to better recognition accuracy.
From each of these estimates, we calculate the respective diffuseness values as \cite{lnt2015-17, galdo_diffuse_2012}:
\begin{equation}
  \setcounter{equation}{17}
  \mathrm{D}(l,f) = \frac{1}{(1 + \widehat{\mathrm{CDR}}(l,f))}.
  \label{eq:diffuseness}
\end{equation}
Subsequently, we take the arithmetic average of all microphone pair-specific diffuseness values, and calculate the final \ac{CDR} estimate as
\begin{equation}
  \widehat{\mathrm{CDR}}_\mathrm{In}(l,f) = \frac{1 - \overline{\mathrm{D}}(l,f)}{\overline{\mathrm{D}}(l,f)},
  \label{eq:CDRestimate_input}
\end{equation}
where $\widehat{\mathrm{CDR}}_\mathrm{In}(l,f)$ describes the final \ac{CDR} estimate at the input of the system, and $\overline{\mathrm{D}}(l,f)$ denotes the average diffuseness obtained by calculating the mean of all microphone pair-specific diffuseness values.
Second, note that the obtained \ac{CDR} estimate $\widehat{\mathrm{CDR}}_\mathrm{In}(l,f)$ is an estimate of the \ac{CDR} at the input of the signal enhancement system, i.e., the beamformer. However, what we actually need is the \ac{CDR} at the output of the beamformer. This can be obtained by applying a correction factor $A_\mathrm{\Gamma}(l,f)$ to $\widehat{\mathrm{CDR}}_\mathrm{In}(l,f)$. Thus, the \ac{CDR} estimate at the output of the beamformer $\widehat{\mathrm{CDR}}_\mathrm{BF}(l,f)$ is defined as
\begin{equation}
  \widehat{\mathrm{CDR}}_\mathrm{BF}(l,f) = \frac{\widehat{\mathrm{CDR}}_\mathrm{In}(l,f)}{ A_\mathrm{\Gamma}(l,f) },
  \label{eq:CDR_bf}
\end{equation}
where $A_\mathrm{\Gamma}(l,f)$ is given as \cite{simmer_post-filtering_2001}
\begin{equation}
  A_\mathrm{\Gamma}(l,f) = \bb{w}^{H}(l,f) \mathbf{J}_\mathrm{diff}(f) \bb{w}(l,f),
\end{equation}
where $\mathbf{J}_\mathrm{diff}(f)$ is the spatial coherence matrix of a diffuse noise field.

\begin{figure}[t]
  \centering
  \scriptsize
  \psfrag{X1}[cl][cl]{$X_{0}(l,f)$}
  \psfrag{XN}[cl][cl]{$X_{N-1}(l,f)$}
  \psfrag{Y_bf}[c][c]{$Y_\mathrm{BF}(l,f)$}
  \psfrag{Y}[cl][cl]{$Y(l,f)$}
  \psfrag{G}[c][c]{\parbox{3cm}{\centering Spectral enhancement\\$G(l,f)$}}
  \psfrag{CDR_in}[cl][cl]{$\widehat{\mathrm{CDR}}_\mathrm{In}(l,f)$}
  \psfrag{CDR_bfout}[cl][cl]{$\widehat{\mathrm{CDR}}_\mathrm{BF}(l,f)$}
  \psfrag{Ainv}[cl][cl]{$1/A_\mathrm{\Gamma}(l,f)$}
\psfrag{Coh estimation}[c][c]{\parbox{1.5cm}{\centering Coherence estimation}}
  \psfrag{Beamforming}[c][c]{\parbox{2cm}{\centering Beamforming}}
  \psfrag{CDR estimation}[c][c]{\parbox{2cm}{\centering CDR\\ estimation}}
  \psfrag{Gamma}[c][c]{$\hat{\Gamma}_\mathrm{x}(l,f)$}
  \includegraphics[width=0.9\columnwidth]{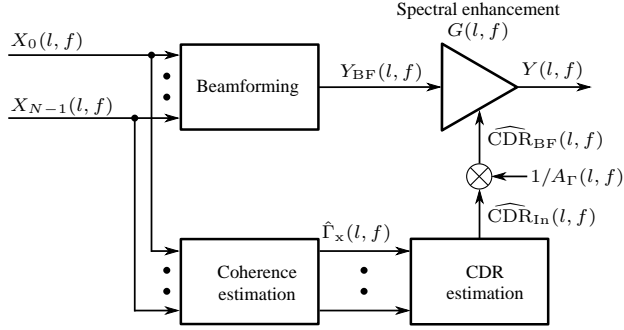}
  \caption{Illustration of the front-end signal processing consisting of beamforming and coherence-based postfilter which is applied to the beamformer output.}
  \label{fig:CDRbased_postfilter}
\end{figure}

%--------------------------------
%Summary of front-end signal processing 
%TODO: genaueres Schaltbild (vor allem mehrere Linien von Kohärenz-Schätzung zu CDR Schätzung)
Fig.~\ref{fig:CDRbased_postfilter} shows the block-diagram of the employed front-end enhancement system, consisting of beamformer and coherence-based postfilter.

%--------------------------------------------------------
%Back-end acoustic modeling
%--------------------------------------------------------
\section{Back-end acoustic modeling}
\label{sec:BackEnd}
As indicated in Fig.~\ref{fig:ASR_pipe}, we employ the acoustic back-end system provided by the \ac{CHiME-3} baseline \ac{ASR} system. It provides an \ac{HMM}-\ac{GMM} system, consisting of $2500$ tied triphone \ac{HMM} states which are modeled by $15000$ Gaussians. The \ac{HMM}-\ac{GMM} system is designed to provide \acp{WER} at relatively low computational costs. In addition, an \ac{HMM}-\ac{DNN} \ac{ASR} system providing state-of-the-art \ac{ASR} performance is contained in the \ac{CHiME-3} baseline. It employs a seven-layer \ac{DNN} with $2048$ neurons per hidden layer and is based on the Kaldi toolkit~\cite{povey_2011}.
%Training
The \ac{DNN} training process includes pre-training using restricted Boltzmann machines, cross entropy training, and sequence discriminative training using the state-level minimum Bayes risk (sMBR) criterion.
For a more detailed presentation of the baseline \ac{ASR} systems, see \cite{Barker_Chime3}.

%--------------------------------------------------------
%Experimental results
%--------------------------------------------------------
\section{Experimental results}
\label{sec:Exper}
In the following, we investigate the impact of our proposed front-end enhancement on the \ac{STFT} spectra of a noisy speech utterance, and evaluate the speech recognition accuracy of the front-end with respect to \acp{WER} using the \ac{CHiME-3} baseline \ac{ASR} systems.

%Setup and parameters
%--------------------------------------------------------
\subsection{Setup and parameters}
%STFT and sampling rate
For all experiments, we use half-overlapping sine windows of $1024$ samples to obtain the complex-valued \ac{STFT} representation of the signals, which is equal to the baseline processing presented in \cite{Barker_Chime3}. The signals were processed at a sampling rate of $16$ kHz.
% Localization
The \ac{DoA} of the desired source, which is required for the \ac{MVDR} beamformer design, was obtained using the baseline localization algorithm \cite{Barker_Chime3}.
% Parameter postfilter
For realizing the coherence-based postfilter, we chose gain floor $G_\mathrm{min}= 0.1$ and overestimation factor $\mu = 1.3$. The short-time coherence estimates $\hat{\Gamma}_\mathrm{x}(l,f)$ were obtained by recursive averaging of the auto- and cross-power spectra with forgetting factor $\lambda = 0.68$, as in \cite{lnt2014-28, lnt2015-17}.

%ASR task
The \ac{ASR} task included sets of real and simulated noisy utterances in four different environments: caf\'e (CAF), street junction (STR), public transport (BUS), and pedestrian area (PED). For each environment, a training set, a development set, and an evaluation set consisting of real and simulated data was provided \cite{Barker_Chime3}.

%Illustration using spectrograms
%--------------------------------------------------------
\subsection{Illustration of front-end impact in the STFT domain}
\begin{figure}
  \subfigure[$20\log_{10}(|S(l,f)|) \, \text{[dB]}$]{
    \hspace{8mm}
    \begin{tikzpicture}[scale=1,trim axis left]
      \begin{axis}[
	label style = {font=\scriptsize},
	tick label style = {font=\tiny},   
	ylabel style={yshift=-1mm},
	xlabel style={yshift=1mm},
	width=8.91cm,height=3.5cm,grid=major,grid style = {dotted,black},  		
	axis on top,
	enlargelimits=false,
	xmin=1, xmax=61, ymin=0, ymax=8000,
	xtick={1,10,20,30,40,50,60},
	xlabel={$l\rightarrow$},
	change y base=true,  y SI prefix=kilo,
	ytick={0, 2000, 4000, 6000, 8000},
	ylabel={$f \, \text{[kHz]}\rightarrow$},
	%colorbar
	colorbar horizontal, colormap/jet, 
	colorbar style={
	  at={(0,1.15)}, anchor=north west, font=\tiny, width=7.325cm, height=0.15cm, xticklabel pos=upper
	},
	point meta min=-120, point meta max=0]
	\addplot graphics [xmin=0, xmax=61, ymin=0, ymax=8000] {spectrogram_S.eps};	
      \end{axis}
      \label{fig:illustration_spectrograms_S}
    \end{tikzpicture}
  }\\[-1mm]       
  \subfigure[$20\log_{10}(|X_{1}(l,f)|) \, \text{[dB]}$]{	
    \hspace{8mm}
    \begin{tikzpicture}[scale=1,trim axis left]
      \begin{axis}[
	label style = {font=\scriptsize},
	tick label style = {font=\tiny}, 
	xlabel style={yshift=1mm},
	ylabel style={yshift=-1mm},  	 
	width=8.91cm,height=3.5cm,grid=major,grid style = {dotted,black},  		
	axis on top, 	
	enlargelimits=false,
	xmin=1, xmax=61, ymin=0, ymax=8000,
	xtick={1,10,20,30,40,50,60},
	xlabel={$l\rightarrow$},
	change y base=true,  y SI prefix=kilo,
	ytick={0, 2000, 4000, 6000, 8000},	
	ylabel={$f \, \text{[kHz]}\rightarrow$}]
	\addplot graphics [xmin=0, xmax=61, ymin=0, ymax=8000] {spectrogram_X1.eps};
      \end{axis}
      \label{fig:illustration_spectrograms_X}
    \end{tikzpicture} 
  }\\[-1mm]
  \subfigure[$20\log_{10}(|Y_\mathrm{BF}(l,f)|) \, \text{[dB]}$]{	
    \hspace{8mm}
    \begin{tikzpicture}[scale=1,trim axis left]
      \begin{axis}[
	label style = {font=\scriptsize},
	tick label style = {font=\tiny}, 
	xlabel style={yshift=1mm},
	ylabel style={yshift=-1mm},  	 
	width=8.91cm,height=3.5cm,grid=major,grid style = {dotted,black},  		
	axis on top, 	
	enlargelimits=false,
	xmin=1, xmax=61, ymin=0, ymax=8000,
	xtick={1,10,20,30,40,50,60},
	xlabel={$l\rightarrow$},
	change y base=true,  y SI prefix=kilo,
	ytick={0, 2000, 4000, 6000, 8000},	
	ylabel={$f \, \text{[kHz]}\rightarrow$}]
	\addplot graphics [xmin=0, xmax=61, ymin=0, ymax=8000] {spectrogram_Y_MVDR.eps};
      \end{axis}
      \label{fig:illustration_spectrograms_YMVDR}
    \end{tikzpicture} 
  }\\[-1mm]  
    \subfigure[$20\log_{10}(|Y(l,f)|) \, \text{[dB]}$]{	
    \hspace{8mm}
    \begin{tikzpicture}[scale=1,trim axis left]
%     \node at (-0.9,1.575) {\scriptsize b)};
      \begin{axis}[
	label style = {font=\scriptsize},
	tick label style = {font=\tiny},
	xlabel style={yshift=1mm},
	ylabel style={yshift=-1mm},  	 
	width=8.91cm,height=3.5cm,grid=major,grid style = {dotted,black},  		
	axis on top, 	
	enlargelimits=false,
	xmin=1, xmax=61, ymin=0, ymax=8000,
	xtick={1,10,20,30,40,50,60},
	xlabel={$l\rightarrow$},
	change y base=true,  y SI prefix=kilo,
	ytick={0, 2000, 4000, 6000, 8000},	
	ylabel={$f \, \text{[kHz]}\rightarrow$}]
	\addplot graphics [xmin=0, xmax=61, ymin=0, ymax=8000] {spectrogram_Y_MVDR_wf.eps};
      \end{axis}
      \label{fig:illustration_spectrograms_Y}
    \end{tikzpicture} 
  }\\[-3mm]
  \subfigure[$D_\mathrm{BF}(l,f)$]{
    \hspace{8mm}
    \begin{tikzpicture}[scale=1,trim axis left]    
      \begin{axis}[
	label style = {font=\scriptsize},
	tick label style = {font=\tiny},   
	xlabel style={yshift=1mm},
	ylabel style={yshift=-1mm},
	width=8.91cm,height=3.5cm,grid=major,grid style = {dotted,black},  		
	axis on top, 	
	enlargelimits=false,
	xmin=1, xmax=61, ymin=0, ymax=8000,	
	xtick={1,10,20,30,40,50,60},
	xlabel={$l\rightarrow$},
	change y base=true,  y SI prefix=kilo,
	ytick={0, 2000, 4000, 6000, 8000},
	ylabel={$f\text{[kHz]}\rightarrow$},
	%colorbar
	colorbar horizontal, colormap/jet, 
	colorbar style={
	  at={(0,1.15)}, anchor=north west, font=\tiny, width=7.325cm, height=0.15cm, xticklabel pos=upper
	},
	point meta min=0, point meta max=1]
	\addplot graphics [xmin=0, xmax=61, ymin=0, ymax=8000] {diffuseness_MVDR.eps};
      \end{axis}
      \label{fig:illustration_spectrograms_D}
    \end{tikzpicture}
    }
  % \vspace{-7mm}
  \caption{Illustration of impact of front-end signal processing on the recorded noisy microphone signal, with recorded close-talking desired signal $S(l,f)$ in (a), microphone signal $X_{1}(l,f)$ in (b), baseline beamformer output signal $Y_\mathrm{BF}(l,f)$ in (c), and postfilter output signal $Y(l,f)$ in (d). Fig. (e) shows the diffuseness $D_\mathrm{BF}(l,f)$ which was estimated from the beamformer output signal in (c), and which has been used to compute the postfilter gains.}
  \label{fig:illustration_spectrograms}
  % \vspace{-4mm}
\end{figure}
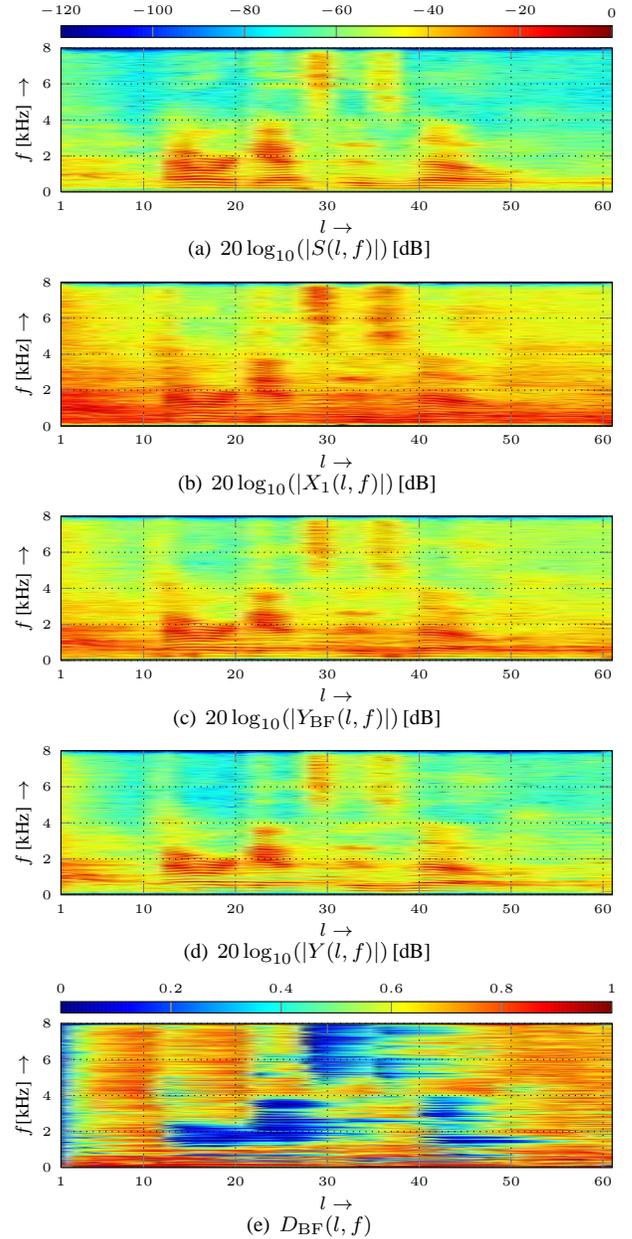

In Fig.~\ref{fig:illustration_spectrograms}, we illustrate the impact of the \ac{MVDR} beamformer and the coherence-based postfilter on the \ac{STFT} spectra of a noisy utterance, with the number of frames $l$ and frequency $f$ on the horizontal and vertical axis, respectively. Note that the coarse temporal resolution of the \ac{STFT} spectra is due to the baseline block-processing. As a reference, the spectrum of the close-talking microphone (channel $0$) is shown in Fig.~\ref{fig:illustration_spectrograms_S}. It contains the desired utterance plus little background noise. The recorded desired signal is a male speaker saying ``\textit{Our guess is no}'' in the caf\'e environment. The spectrum of microphone $1$ is illustrated in Fig.~\ref{fig:illustration_spectrograms_X}. As can be seen, low- as well as high-frequency noise is acquired by the microphone, whereas most of the noise is present in the frequency range of speech.
% MVDR
Applying the baseline \ac{MVDR} beamformer leads to a reduction of the interfering components, as illustrated in Fig.~\ref{fig:illustration_spectrograms_YMVDR}.
%Postfilter
A comparison of Fig.~\ref{fig:illustration_spectrograms_YMVDR} with Fig.~\ref{fig:illustration_spectrograms_Y} shows that applying the coherence-based postfilter to the \ac{MVDR} beamformer output yields a significant reduction of interference across the entire frequency range, but it also removes low-frequency components of the desired signal.
The estimated diffuseness $D_\mathrm{BF}(l,f)$ at the beamformer output is illustrated in Fig.~\ref{fig:illustration_spectrograms_D}. Comparing Figs.~\ref{fig:illustration_spectrograms_D} and \ref{fig:illustration_spectrograms_YMVDR} shows that $D_\mathrm{BF}(l,f)$ is very low whenever the desired source is active, which is to be expected, since the \ac{CDR} will be high whenever the desired source is active.
A final comparison of Figs.~\ref{fig:illustration_spectrograms_S} and \ref{fig:illustration_spectrograms_Y} reveals the similarity between the front-end output signal $Y(l,f)$ and the close-talking microphone signal $S(l,f)$, which indicates the effectiveness of the proposed front-end signal enhancement technique.

%Evaluation with respect to WER
%--------------------------------------------------------
\subsection{Evaluation of estimation accuracy}
% In the following, the \acp{WER} of the baseline and the extended front-end obtained with the baseline \ac{ASR} system are presented and compared. 

%--------------------
%average WERs for HMM-GMM and HMM-DNN baseline ASR systems
%--------------------
\begin{table*}[t]  
  \begin{center}    
    \caption{Average \acp{WER} (in \%) obtained with the baseline (MVDR) and extended (MVDR+PF) front-end signal enhancement for the baseline \ac{HMM}-\ac{GMM} and \ac{HMM}-\ac{DNN} \ac{ASR} systems.}
    \label{tab:Results_avgWERs}   
    \begin{tabular}{c c c c c c c}
      \toprule   
      \multirow{2}{2cm}{\centering \raisebox{-1mm}[0pt]{Acoustic model}} & \multirow{2}{2cm}{\centering \raisebox{-1mm}[0pt]{Test data}} & \multirow{2}{2cm}{\centering \raisebox{-1mm}[0pt]{Training data}} & \multicolumn{2}{c}{Development set} & \multicolumn{2}{c}{Evaluation set}\\
      \cmidrule(r){4-7}
      & & & Real data & Sim. data & Real data & Sim. data\\
      \midrule
      %noisy
      HMM-GMM & \multirow{2}{2cm}{\centering Noisy} & \multirow{2}{2cm}{\centering Noisy} & 18.67 & 18.07 & 32.97 & 21.89 \\    
      HMM-DNN+sMBR &  &  & 16.70 & 14.38 & 34.53 & 21.34 \\ %eigene Ergebnisse
      \midrule
      %MVDR
      HMM-GMM & \multirow{2}{2cm}{\centering MVDR} & \multirow{2}{2cm}{\centering MVDR} & 20.87 & 9.67 & 38.18 & 10.99 \\    
  %     HMM-DNN &  &  & 21.06 & 9.20 & 43.47 & 12.09 \\
      HMM-DNN+sMBR &  &  & 17.70 & 8.22 & 33.88 & 10.79 \\
      \midrule
      %MVDR+PF
      HMM-GMM & \multirow{2}{2cm}{\centering MVDR+PF} & \multirow{2}{2cm}{\centering MVDR+PF} & 16.13& 11.55 & 28.29 & 12.87 \\    
  %     HMM-DNN & & & xx.xx & xx.xx & xx.xx & xx.xx \\
      HMM-DNN+sMBR & & & 14.97 & 10.17 & 28.68 & 15.24 \\
	\bottomrule    
    \end{tabular}
  \end{center}
  \begin{center}
    \caption{\acp{WER} (in \%) obtained with the extended front-end signal enhancement for the baseline \ac{HMM}-\ac{DNN} \ac{ASR} system in each scenario.}
    \label{tab:Results_MVDR+PF_scenarioSpecificWERs}
    \begin{tabular}{c c c c c}
      \toprule   
      \multirow{2}{2cm}{\centering \raisebox{-1mm}[0pt]{Environment}} & \multicolumn{2}{c}{Development set} & \multicolumn{2}{c}{Evaluation set}\\
      \cmidrule(r){2-5}
      & Real data & Sim. data & Real data & Sim. data\\
      \midrule
      BUS & 17.63 & 8.94 & 35.58 & 11.52 \\
      CAF & 14.65 & 12.23 & 32.69 & 17.37 \\
      PED & 12.97 & 8.42 & 26.61 & 15.48 \\
      STR & 14.64 & 11.11 & 19.85 & 16.57 \\
      \bottomrule    
    \end{tabular}
  \end{center}
\end{table*}
Table~\ref{tab:Results_avgWERs} summarizes the average \acp{WER} (in \%) of the baseline (MVDR) and the extended (MVDR+PF) front-end enhancement obtained for the \ac{CHiME-3} baseline \ac{HMM}-\ac{GMM} and \ac{HMM}-\ac{DNN} \ac{ASR} (termed HMM-DNN+sMBR in the tables to be consistent with \cite{Barker_Chime3}) systems. The \acp{WER} were averaged over all four acoustic environments. In the first column the employed acoustic model is specified. The test and training data sets are indicated in the second and third column, whereas the respective results for the development and evaluation data set are given in the fourth and fifth column. The \ac{ASR} systems have always been trained on the output signals of the applied front-end enhancement. As a reference, the first row in Table~\ref{tab:Results_avgWERs} contains the \acp{WER} obtained for the noisy unprocessed microphone signals. Note that the results in the case of no front-end enhancement (Noisy) and for the baseline \ac{MVDR} beamformer (second row in Table~\ref{tab:Results_avgWERs}) only differ slightly from the presented results in \cite{Barker_Chime3}. The slight deviations are due to random initialisation and machine-specific issues.

%Noisy vs MVDR
%Vergleich unprocessed mit MVDR und MVDR+PF
%HMM-GMM
When comparing the results of the \ac{HMM}-\ac{GMM} \ac{ASR} system in the first and second row, one can observe that the baseline front-end enhancement only improves the \acp{WER} for simulated data. In the case of real data, the recognition accuracy of the baseline front-end processing is significantly worse than without front-end signal processing. 
%HMM-DNN
For the \ac{HMM}-\ac{DNN}-based recognizer, significant \ac{WER} improvements can be observed for simulated data, whereas for real data there is no clear advantage of the baseline front-end processing compared to no front-end processing.
 
%MVDR vs MVDR+PF
%HMM-GMM
A comparison of the results for the \ac{HMM}-\ac{GMM} \ac{ASR} system in the second and third row shows that applying the coherence-based postfilter to the \ac{MVDR} beamformer output signal drastically decreases the average \ac{WER} for real data with an improvement of $4.74$ and $9.89$ percentage points for the development and evaluation data set, respectively.
It can also be seen that the \acp{WER} of the extended front-end are slightly increased for simulated data. The reason for this may be that the employed postfilter parameters $\mu$ and $G_\mathrm{min}$ are suboptimal for the simulated data set. 
%
%HMM-DNN
The results for the baseline (MVDR) and the proposed front-end (MVDR+PF) obtained with \ac{HMM}-\ac{DNN} \ac{ASR} system in the second and third row show  the same tendencies. Our proposed front-end enhancement yields significantly lower \acp{WER} for real data and a worse recognition accuracy for simulated data. In the case of real data, the \acp{WER} were decreased by $2.73$ and $5.2$ percentage points for the development and evaluation data set, respectively, by applying the coherence-based postfilter.

% 
%MVDR+PF - interner Vergleich
It is interesting to note that for our proposed front-end, the \ac{HMM}-\ac{DNN} \ac{ASR} system only yields a better recognition performance than the \ac{HMM}-\ac{GMM} system for the development data, whereas for the real evaluation data the \ac{HMM}-\ac{GMM} \ac{ASR} system achieves lower \acp{WER}. Especially for the simulated evaluation data, the \ac{HMM}-\ac{GMM} \ac{ASR} is superior to the \ac{HMM}-\ac{DNN}-based recognizer. 
One explanation for this phenomenon might be a suboptimal architecture of the \ac{DNN} which we did not optimize as part of this contribution.
Finally, we can observe that only applying the postfilter to the \ac{MVDR} output signal yields significantly lower \acp{WER} with both baseline \ac{ASR} systems for real data, compared to the unprocessed signal, which confirms the effectiveness of our proposed postfilter.

%--------------------
%Scenario-specific WERs for HMM-DNN baseline ASR system
%--------------------
In Table~\ref{tab:Results_MVDR+PF_scenarioSpecificWERs} the scenario-specific \acp{WER} of our proposed front-end enhancement obtained with the baseline \ac{HMM}-\ac{DNN} \ac{ASR} system are provided. Judging from the obtained \acp{WER}, the BUS environments seems to be the most challenging scenario for real data, whereas the highest \ac{WER} for simulated data was obtained for the caf\'e scenario.

%%%%%%%%%%%%%%%%%%%%%%%%%%%%%%%%%%%%%%%%
% Conclusion
%%%%%%%%%%%%%%%%%%%%%%%%%%%%%%%%%%%%%%%%
\section{Conclusion}
\label{sec:summary_conclusion}
In this contribution to the \ac{CHiME-3} challenge, we proposed an extension of the baseline front-end speech enhancement by a coherence-based postfilter. The postfilter is realized as a Wiener filter, where an estimate of the ratio between direct and diffuse signal components at the output of the baseline \ac{MVDR} beamformer is used as an approximation of the short-time \ac{SNR} to compute the filter gains. To estimate the ratio between direct and diffuse signal components, we used a \ac{DoA}-independent estimator, which can be efficiently realized since it only requires an estimate of the auto- and cross-power spectra at the microphone signals. As a consequence, the proposed postfilter has a very low computational complexity as well.
Both the baseline and the extended front-end speech enhancement have been evaluated on real and simulated data with respect to \acp{WER} using the baseline \ac{HMM}-\ac{GMM} and \ac{HMM}-\ac{DNN} \ac{ASR} systems. The results    confirmed that the proposed coherence-based postfilter significantly improves the recognition accuracy of the enhanced speech compared to the \ac{MVDR} beamformer when applied to real data. The improved recognition accuracy in addition to the low computational complexity makes the proposed postfilter very suitable for real-time robust distant speech recognition.  
Future work includes the analysis of the performance of \ac{DoA}-dependent \ac{CDR} estimators for the \ac{CHiME-3} data. Also combining \ac{DoA}-dependent and \ac{DoA}-independent \ac{CDR} estimators in different frequency ranges will be investigated. Moreover, using spatial diffuseness features as an additional input to a DNN-based acoustic model, as proposed in \cite{lnt2015-1}, is another avenue for future work.

%%%%%%%%%%%%%%%%%%%%%%%%%%%%%%%%%%%%%%%%
% Acknowledgements
%%%%%%%%%%%%%%%%%%%%%%%%%%%%%%%%%%%%%%%%
\section{Acknowledgement}
\label{sec:acknowledgements}
We would like to thank Stefan Meier and Christian Hofmann for their continuous support and fruitful discussions.

%%%%%%%%%%%%%%%%%%%%%%%%%%%%%%%%%%%%%%%%
% Bibliography
%%%%%%%%%%%%%%%%%%%%%%%%%%%%%%%%%%%%%%%%
\newpage
% \bibliographystyle{IEEEbib}
% \bibliography{chime2015}

\end{document}